\documentclass[english,aps]{revtex4}
\usepackage[latin9]{inputenc}
\usepackage{url}
\usepackage{natbib}
\newcommand{\doilink}[1]{doi:\href{http://dx.doi.org/#1}{#1}}
\usepackage{textcomp}
\usepackage{amsmath}
\usepackage{amssymb}
\usepackage{graphicx}
\usepackage[unicode=true,
 bookmarks=true,bookmarksnumbered=false,bookmarksopen=false,
 breaklinks=false,pdfborder={0 0 1},backref=false,colorlinks=false]
 {hyperref}
\hypersetup{pdftitle={A quantum cascade laser-based mid-IR frequency metrology system with ultra-narrow linewidth and 1*10\^-13 -level frequency instability},
 pdfauthor={Michael G. Hansen, Evangelos Magoulakis, Qun-Feng Chen, Ingo Ernsting, Stephan Schiller}}

\makeatletter
\@ifundefined{textcolor}{}
{%
 \definecolor{BLACK}{gray}{0}
 \definecolor{WHITE}{gray}{1}
 \definecolor{RED}{rgb}{1,0,0}
 \definecolor{GREEN}{rgb}{0,1,0}
 \definecolor{BLUE}{rgb}{0,0,1}
 \definecolor{CYAN}{cmyk}{1,0,0,0}
 \definecolor{MAGENTA}{cmyk}{0,1,0,0}
 \definecolor{YELLOW}{cmyk}{0,0,1,0}
}

\makeatother

\begin{document}

\title{A quantum cascade laser-based mid-IR frequency metrology system with
ultra-narrow linewidth and $1\times10^{-13}$-level frequency instability}

\author{Michael G. Hansen}

\email{michael.hansen@uni-duesseldorf.de}

\author{Evangelos Magoulakis}

\author{Qun-Feng Chen}

\author{Ingo Ernsting}

\author{Stephan Schiller}

\affiliation{Institut f\"ur Experimentalphysik, Heinrich-Heine-Universit\"at D\"usseldorf,
40225 D\"usseldorf, Germany}
\begin{abstract}
We demonstrate a powerful tool  for high-resolution mid-IR spectroscopy
and frequency metrology with quantum cascade lasers (QCLs). We have
implemented frequency stabilization of a QCL to an ultra-low expansion
(ULE) reference cavity, via upconversion to the near-IR spectral range,
at a level of $1\times10^{-13}$. The absolute frequency of the QCL
is measured relative to a hydrogen maser, with instability $<1\times10^{-13}$
and inaccuracy $5\times10^{-13}$, using a frequency comb phase-stabilized
to an independent ultrastable laser. The QCL linewidth is determined
to be 60~Hz, dominated by fiber noise. Active suppression of fiber
noise could result in sub-10~Hz linewidth.
\end{abstract}

%\pacs{(120.0120) Instrumentation, measurement, and metrology; (190.0190) Nonlinear optics; (140.5965) Semiconductor lasers, quantum cascade; (300.6390) Spectroscopy, molecular.}

\maketitle
High-resolution spectroscopy in the mid-IR has been a crucial tool
for the elucidation of fine details of molecular structure and dynamics
for many decades. Earlier, mostly gas lasers (CO$_{2}$, CO, He-Ne)
with narrow linewidth (kHz to sub-Hz, e.g.\ \citep{Bernard1997}) have
been used as sources, permitting spectral resolutions below the Doppler
width of gaseous room temperature samples, by employing appropriate
nonlinear spectroscopic techniques. The molecular gas lasers are only
line-tunable, in some cases restricting severely their applicability.
As a remedy, the generation of tunable microwave sidebands has been
successfully implemented \citep{Meyer1995}, but the method has the
drawback of the low power in the sidebands. Another approach is the
difference-frequency generation (DFG) from near-IR sources which allows
a wide tuning range, but generally suffers from low mid-IR power \citep{Wang2011}. 

Quantum cascade lasers (QCLs) offer both high power and relatively
wide tuning. The free-running linewidth on the timescale of 1~s is
typically on the order of 1~MHz and has been the subject of several
studies \citep{Bartalini2011,Tombez2012}. Linewidth narrowing, alongside
absolute frequency stabilization and finally absolute frequency measurement,
are therefore important tasks for rendering QCLs usable for high-resolution
spectroscopy.

It has already been shown that the QCL linewidth can be dramatically
reduced, by locking to a reference cavity \citep{Taubman2004}, by
phase-locking to a narrow-linewidth CO$_{2}$ reference laser, reaching
less than 10~Hz relative linewidth \citep{Sow2014}, or by phase-locking
to the DFG wave generated from two near-IR cw lasers \citep{Galli2013}.
This latter approach however limits the spectral coverage of the QCL,
since it relies on particular reference lasers. 

The upconversion approach \citep{Amy-Klein2005} appears as particularly
suitable as a general approach for QCL frequency metrology and linewidth
narrowing \citep{Mills2012}. References~\citep{Mills2012,Galli2013}
also performed absolute frequency measurements. In our own work \citep{Hansen2013},
we upconverted radiation from a 5.4~$\mu$m QCL to 1.2~$\mu$m by
mixing with a cw 1.5~$\mu$m fiber laser. These two near-IR waves
were simultaneously measured by and stabilized to a standard Er:fiber
comb referenced to a hydrogen maser. No linewidth narrowing was implemented
at the time.

In the present work, we extend significantly the performance of QCL
frequency metrology, by demonstrating absolute frequency stabilization,
linewidth narrowing, and absolute frequency measurement at the 10~Hz
level without relying on an ultrastable mid-IR reference laser. An
important advantage of the present approach is that it is applicable
to any QCL  in the range 5 - 12~\textmu m. 

In practice, a flexible solution for both linewidth-narrowing and
absolute frequency stabilization is provided by locking the QCL to
an ultra-stable reference cavity made of ultra-low expansion glass
(ULE). This approach is a common one for lasers emitting in the near-IR
and visible spectral ranges. However, to our knowledge, it has not
yet been implemented with QCLs. The use of a reference resonator has
the advantage that the QCL is stabilized without the need of a frequency
comb, a significant simplification. For applications where the optical
frequency of the QCL must also be measured or accurately monitored,
a comb is nevertheless still required. For such cases, we propose
and demonstrate here an approach where not the QCL but rather the
upconverted radiation is stabilized to a ULE reference cavity. This
has the significant advantage that the reference cavity can use coatings
for the near-IR and that less expensive and higher-performance near-IR
rather than mid-IR optical components can be employed.

\begin{figure*}[t]
\includegraphics[width=16cm]{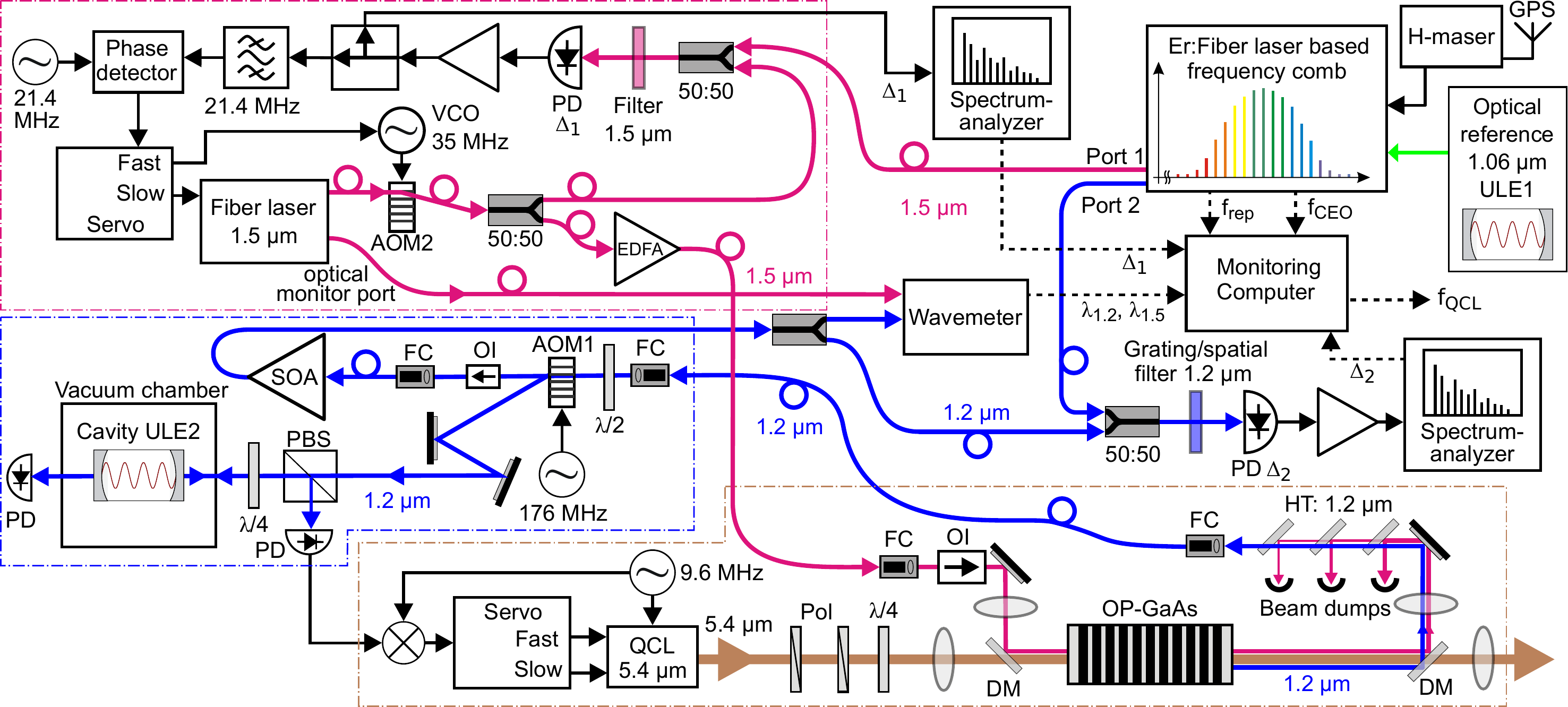}

\protect\caption{\label{fig:setup}Schematic of the setup. Pink: fiber laser wave
at 1.5~$\mu$m, brown: QCL wave at 5.4~$\mu$m, blue: sum frequency
wave at 1.2~$\mu$m. Full lines: analog electric signals; dashed
lines: digital signals. DM: dichroic mirror; AOM: acousto-optic modulator;
SOA: semiconductor optical amplifier; Pol: wire grid polarizer; PD:
photodetector; OI: optical Faraday isolator; FC: fiber collimator;
PBS: polarizing beam splitter; VCO: voltage-controlled oscillator.
The spectrum analyzers are referenced to the H-maser.}
\end{figure*}

The basic concept of the spectrometer for mid-IR frequency metrology
has been described before \citep{Hansen2013}. In this work, the spectrometer
has been further developed such that it provides fast frequency stabilization
of a QCL and significant linewidth narrowing. The overall concept
is as follows. A standard near-IR fiber-based frequency comb (pulse
repetition frequency $f_{rep}$) is phase-locked to a ULE cavity (ULE1)
stabilized laser. This provides a comb of ultra-narrow modes with
Hz-level linewidths. A near-IR laser (frequency $f_{1.5}$) is phase-locked
to the nearest mode of the frequency comb. The sum-frequency (upconverted)
wave of the near-IR laser and of the QCL (frequency $f_{1.2}=f_{1.5}+f_{QCL}$),
is frequency-stabilized to another stable cavity (ULE2), by controlling
$f_{QCL}$. Therefore, the QCL frequency $f_{QCL}$ is also stabilized.
Furthermore, the frequencies $f_{1.5}$, $f_{1.2}$ and $f_{rep}$
are simultaneously measured by the comb and a hydrogen maser (H-maser)
in real time to allow measuring $f_{QCL}$ and its spectral properties,
i.e.~its fluctuations.

A schematic of the apparatus in shown in Fig.~\ref{fig:setup}. An
erbium-doped fiber-based frequency comb is phase-locked to a 1064~nm Nd:YAG
laser which is stabilized to a compact and robust reference cavity
(ULE1), achieving $1\times10^{-15}$ short-term frequency instability
\citep{Chen2014}. The phase-lock acts on the cavity length of the
frequency comb via an intracavity electrooptic phase modulator and
a piezo-actuated cavity mirror, controlling the repetition rate such
that the beat ($\Delta_{ULE1}$) between the Nd:YAG laser (frequency
$f_{ULE1}$) and a nearby comb mode (integer mode number $m_{ULE1}$)
has a fixed value, $\Delta_{ULE1}$. The residual linewidth of this
beat is less than 1~Hz. The comb repetition rate $f_{rep}$ is then
determined by the frequency of the optical reference ULE1 through
the phase-lock condition $m_{ULE1}f_{rep}\pm f_{CEO}\pm\Delta_{ULE1}=f_{ULE1}$.
The carrier envelope offset frequency $f_{CEO}$ is stabilized to
a fixed-frequency RF signal (20~MHz). In this way, all modes of the
comb are of high absolute stability and ultra-narrow. Using a separate
ultra-narrow 1.5~$\mu$m laser, we determined an upper limit of 6.5~Hz
(resolution bandwidth limited) for the linewidth of comb modes at
this wavelength. 

The repetition rate $f_{rep}$ is measured by a frequency counter
referenced to the H-maser, with 1~s integration time. The frequency
instability of the maser is below $2\times10^{-14}$ for integration
times longer than 10~s; its frequency is determined to an uncertainty
$5\times10^{-13}$ by comparison with GPS signals. The values of $f_{rep}$,
$f_{CEO}$, combined with the frequency $\Delta>0$ of the beat of
a given laser wave (frequency $f$) and a comb mode (mode number $m$),
yield the actual optical frequency, $f=m\, f_{rep}\pm f_{CEO}\pm\Delta$,
where $m$ and the sign associated with $\Delta$ are determined using
a wavemeter having suitable accuracy. The sign associated with $f_{CEO}$
is determined by the corresponding phase-lock circuit.

The employed QCL is a cw room-temperature, distributed feedback grating,
ridge waveguide QCL (Maxion). A two-stage temperature stabilization
system allows the QCL to be operated between -10~\textdegree C and
+13~\textdegree C, corresponding to a tuning range from $\lambda_{QCL}$=
5350~nm to 5368 nm, with an output power of 40~mW. For sum-frequency
generation, its radiation is overlapped with radiation from an Erbium-doped
fiber-amplifier (EDFA), which is seeded by a 1.5~$\mu$m fiber laser
and then sent to an orientation-patterned (OP) GaAs crystal. 30~$\mu$W
of the sum-frequency wave at 1.2~$\mu$m are delivered via a 2~m
long fiber to an actively vibration isolated breadboard carrying a
second ULE cavity (ULE2). An AOM (AOM1) is used to suppress interferences
between the fiber and the ULE cavity, and a circulator guides the
wave reflected from the cavity onto a photodiode for error signal
detection. The ULE cavity spacer is 8.4 cm long and the cavity finesse
is 160\,000. A two-stage temperature stabilization system acts on
the cavity block inside the vacuum chamber and the chamber itself.
The 1.2~$\mu$m radiation is stabilized to the cavity by the Pound-Drever-Hall
technique by controlling the frequency of the QCL. The 1.2~$\mu$m
radiation is frequency modulated by modulating the QCL current at
9.6~MHz through a bias-tee, which is also used for the fast frequency
feedback signal. The slow frequency feedback signal is applied to
the modulation input of the current driver.

The zero-order output of AOM1 is led through an optical isolator and
then coupled into an optical amplifier which boosts the power level
of the 1.2~$\mu$m radiation to approx. 0.1~mW. This wave is then
sent via a 30~m long fiber to the frequency comb, where it produces
a beat $\Delta_{2}$ with a comb mode.

\begin{figure}[t]
\includegraphics[width=8.4cm]{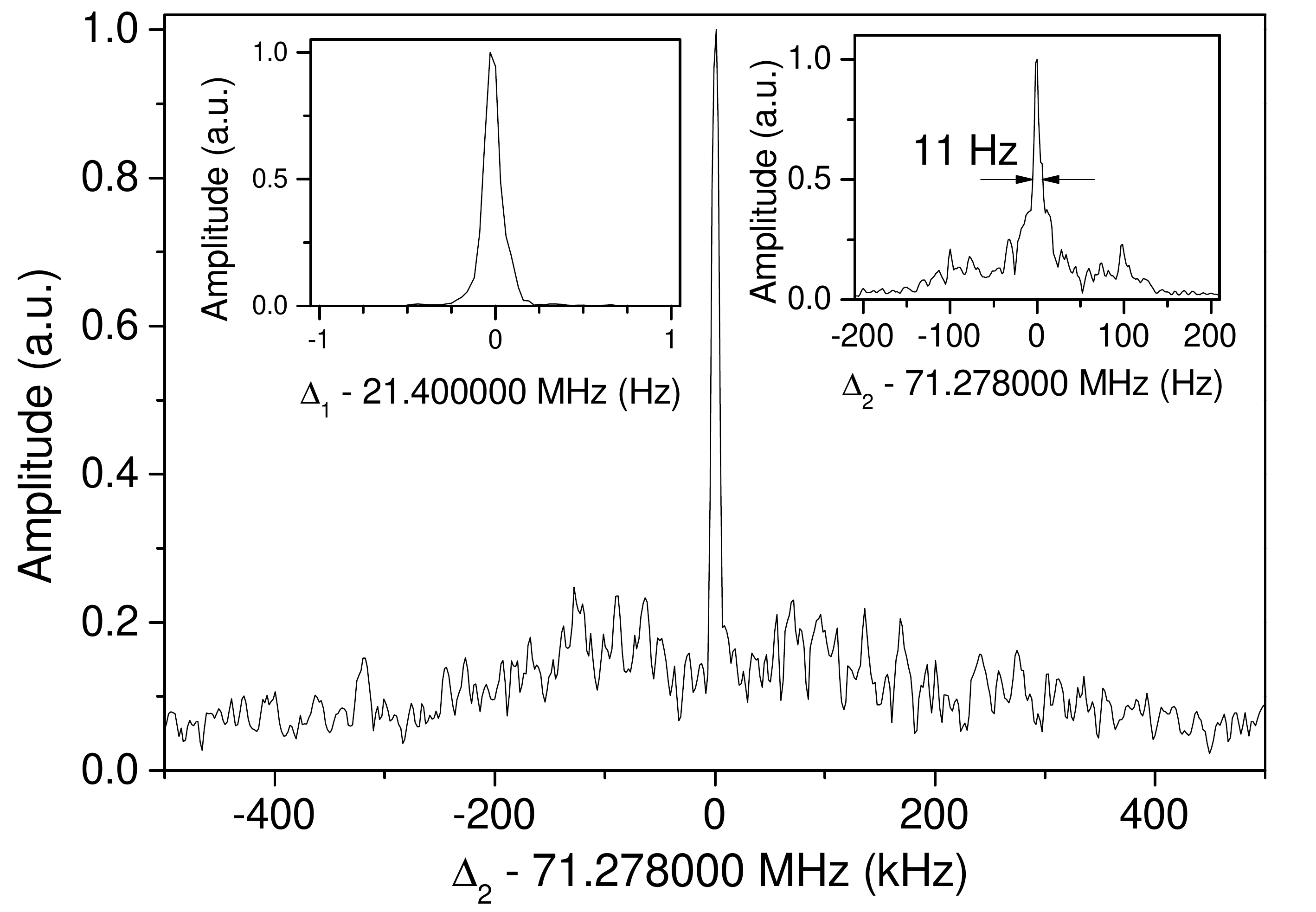}

\protect\caption{\label{fig:beat-qcl}Beat note $\Delta_{2}$ of the 1.2~$\mu$m upconverted
wave with a mode of the frequency comb. The wave is not locked to
this mode. Resolution bandwidth (RBW) 6.4~kHz, sweep time 21~ms.
Left inset: beat note $\Delta_{1}$ of the 1.5~$\mu$m laser with
a mode of the frequency comb. The laser frequency is phase-locked
to this mode, FWHM 0.1~Hz, RBW 31.25~mHz, sweep time 32~s. Right
inset: high-resolution measurement of the 1.2~$\mu$m beat, 11~Hz
FWHM, RBW 6.4~Hz, sweep time 2.2~s. }
\end{figure}

The 1.5~$\mu$m fiber laser is phase-locked to the frequency comb
such that the beat frequency $\Delta_{1}$ of the fiber laser and
the comb is constant by means of an AOM (AOM2) and the piezo inside
the laser. Then, $f_{1.5}=m_{1}f_{rep}\pm f_{CEO}\pm\Delta_{1}$.
The fiber laser is connected to the frequency comb using a second
30~m long fiber.A spectrum of the beat $\Delta_{2}$ between the
frequency comb and the 1.2~$\mu$m wave is shown in Fig.~\ref{fig:beat-qcl},
the FWHM is 11\textbf{~}Hz. The beat $\Delta_{1}$ of the fiber laser
with the comb line to which it is phase-locked is also shown in the
figure; it has sub-Hz linewidth. The two measured linewidths need
to be corrected for the influence of the fiber noise of the two 30~m
long fibers, which are not actively stabilized. The fiber noise was
measured by sending the 1.5~$\text{\ensuremath{\mu}}$m laser wave
in a round-trip through the two fibers and performing a self-beat.
This resulted in 120~Hz FWHM. This implies that at the location of
the 1.5~$\mu$m laser, its linewidth is on the order of 60~Hz. On
the other hand, the linewidth of the 1.2~$\mu$m wave is expected
to be substantially less than 11~Hz, since the 1.2~\textmu m light
used for making a beat with the comb has been broadened by propagation
through the 30~m long fiber. The linewidth of the QCL is not observed
directly, but can be inferred from the geometric sum of the 1.2~\textmu m
and the 1.5~\textmu m linewidths, to be approximately 60~Hz, dominated
by the fiber noise on the 1.5~$\mu$m wave. Active suppression of
the fiber noise could result in linewidth-narrowing of the QCL to
the sub-10~Hz level, i.e. to the inferred linewidth of the 1.2~$\mu$m
wave at the ULE2 cavity. 

The absolute frequency properties of the QCL are determined as follows.
The two near-IR frequencies are given by $f_{1.2}=m_{2}f_{rep}\pm f_{CEO}\pm\Delta_{2}$
and $f_{1.5}=m_{1}f_{rep}\pm f_{CEO}\pm\Delta_{1}$, where $m_{1}$
and $m_{2}$ are the respective constant mode numbers computed from
wavemeter readings, and the values of $\Delta_{1}$ and $\Delta_{2}$
are measured by spectrum analyzers (2~s integration time) every 5~s
via peak detection. The QCL's optical frequency is then computed as
$f_{QCL}=(m_{2}-m_{1})f_{rep}\pm(\Delta_{2}\mp\Delta_{1})$ by the
measurement computer in real time. 

Note that both the upconverted wave's frequency and the QCL's frequency
are essentially determined by the frequencies of the ULE cavities.
While $f_{1.2}$ is directly given by a cavity mode frequency of ULE2
($f_{1.2}=f_{ULE2})$, $f_{QCL}$ is determined by cavity mode frequencies
of both ULE1 and ULE2, according to $f_{QCL}=-(m_{1}/m_{ULE1})\, f_{ULE1}+f_{ULE2}$~$-(m_{1}/m_{ULE1})\,(f_{CEO}\mp\Delta_{ULE1})$~$-f_{CEO}\mp\Delta_{1}$.
We recall that $f_{CEO},\,\Delta_{ULE1},\,\Delta_{1}$ are constant
RF frequencies. Therefore, the frequencies $f_{1.2}$ and $f_{QCL}$
will be affected by the (small) drifts of the ULE cavities' frequencies
due to temperature variations and by mechanical relaxation, as well as
by locking errors.

\begin{figure}[t]
\includegraphics[width=8.4cm]{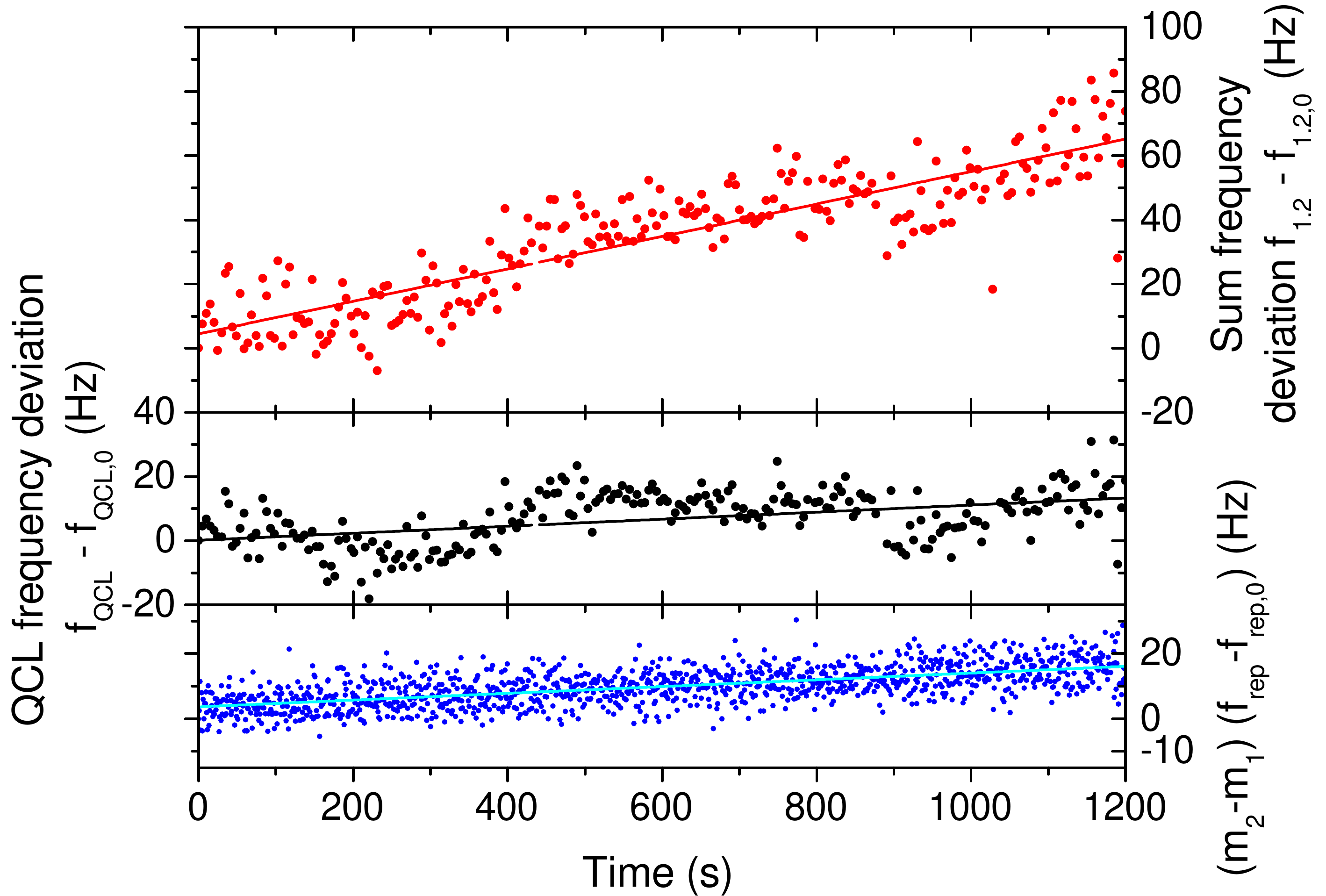}

\protect\caption{\label{fig:qcl_longterm}Simultaneous frequency measurement of the
scaled comb repetition rate $(m_{2}-m_{1})f_{rep}$, of the upconverted
wave's $f_{1.2}$, and of the QCL's $f_{QCL}$, respectively. Lines:
Linear fits. $(m_{1}-m_{2})f_{rep,0}=$ $55.916\,742\,336\,214(5)\,{\rm THz}$.}
\end{figure}

A measurement of the scaled comb repetition rate $(m_{2}-m_{1})f_{rep}$
as a function of time is shown in Fig.~\ref{fig:qcl_longterm}. Its
drift is due to the drift of ULE1, which we observed to slowly vary
with time. The frequency instability, due to counter noise and H-maser
frequency instability, is 5~Hz on the 1~s time scale, and averages
down to $\le1$~Hz at integration times $>10$~s (after drift removal).

The upconverted wave's frequency and QCL's frequency are also presented
in Fig.~\ref{fig:qcl_longterm}. The frequencies $f_{1,2},\, f_{QCL}$
have small drifts of 0.05~Hz/s and 0.01~Hz/s, respectively. These
drifts are not constant in time but are typically at this level and
may be attributed to the reference cavities ULE1 and ULE2. We emphasize
that the QCL frequency is measured in real time and therefore in an
actual spectroscopic experiment it will be possible to correct for
the observed drift either by active control, or during the data analysis.
The frequency instability of the QCL is 5~Hz ($1\times10^{-13}$
in fractional terms) for averaging times larger than 10~s (after
drift removal). The small systematic deviations of the QCL frequency
are of order 30~Hz; the above determination of the instability of
$(m_{2}-m_{1})f_{rep}$ and the negligible instability of the beat
$\Delta_{1}$ indicates that the deviations are not dominantly due
to the comb and fiber laser but due to residual errors of the QCL
lock to cavity ULE2, i.e. they are expressed via the $\Delta_{2}$
contribution to $f_{QCL}$. The absolute optical frequencies at $t=0\,$s
were measured to be $f_{1.2,0}=247.477\,669\,529\,3(2)\,{\rm THz}$
and $f_{QCL,0}=55.916\,762\text{\,}922\,53(3)\,{\rm THz}$. The uncertainties
are due to the H-maser frequency uncertainty.

In conclusion, we have demonstrated both absolute frequency stabilization
at a level of $1\times10^{-13}$ (after linear drift removal) and
linewidth narrowing of a standard QCL to the 60~Hz level ($1\times10^{-12}$).
Absolute frequency measurement is also achieved, with uncertainty
of 30~Hz ($5\times10^{-13})$. The two key ingredients are a high-finesse
ULE cavity for the near-IR and an optically referenced near-IR frequency
comb. The present system appears very well suited for ultra-high resolution
spectroscopy, e.g. of cold, trapped molecules. The QCL linewidth is
broadened by fiber noise, but this may be eliminated in a straightforward
manner in future work and would result in sub-10~Hz linewidth.

The described approach has the advantage that it combines strong linewidth
narrowing and absolute frequency stabilization with absolute frequency
measurement. The approach is general, since the employed nonlinear
crystal (GaAs) is suitable for upconversion of QCL wavelengths covering
the whole mid-IR. However, alternative crystals could also be used.
We note that a first convenient simplification would consist in stabilizing
the 1.5~$\mu$m laser to a ULE cavity and using the laser both for
upconversion of the QCL and as a reference laser for stabilization
of the comb. A second simplification is to use only a single near-IR
ULE cavity, with appropriately designed mirror coatings, instead of
two cavities.

One additional outcome of this work is the capability of observing
the linewidth of the QCL in real-time via the observation of the beat
between the upconverted wave and a comb mode, provided that fiber
noise in the connection between the fiber laser, the generated sum-frequency
wave and the frequency comb is either actively suppressed or small
compared to the observed linewidth. The other sources of linewidth
broadening, upconversion and comb, contribute negligibly to the beat
linewidth.

We thank A.~Nevsky, U.~Rosowski, and E.~Wiens for assistance and
measurement of the comb linewidth. This work was partially supported
by the DFG (project Schi~431/19-1).

%\bibliographystyle{test2}
%\bibliography{references}

\begin{thebibliography}{10}
\providecommand{\url}[1]{\texttt{#1}}
\providecommand{\urlprefix}{URL }
\expandafter\ifx\csname urlstyle\endcsname\relax
  \providecommand{\doi}[1]{doi:\discretionary{}{}{}\doilink{#1}}\else
  \providecommand{\doi}{doi:\discretionary{}{}{}\begingroup
  \urlstyle{rm}\Url}\fi
\providecommand{\eprint}[2][]{\url{#2}}

\bibitem{Amy-Klein2005}
A.~Amy-Klein, A.~Goncharov, M.~Guinet, C.~Daussy, O.~Lopez, A.~Shelkovnikov,
  and C.~Chardonnet.
\newblock \emph{Absolute frequency measurement of a {SF$_6$} two-photon line by
  use of a femtosecond optical comb and sum-frequency generation}.
\newblock \emph{Opt. Lett.}, \textbf{30}(24):3320--3322 (Dec 2005).
\newblock \doilink{10.1364/OL.30.003320}.

\bibitem{Bartalini2011}
S.~Bartalini, S.~Borri, I.~Galli, G.~Giusfredi, D.~Mazzotti, T.~Edamura,
  N.~Akikusa, M.~Yamanishi, and P.~{De Natale}.
\newblock \emph{Measuring frequency noise and intrinsic linewidth of a
  room-temperature {DFB} quantum cascade laser}.
\newblock \emph{Opt. Express}, \textbf{19}(19):17996--18003 (Sep 2011).
\newblock \doilink{10.1364/OE.19.017996}.

\bibitem{Bernard1997}
V.~Bernard, C.~Daussy, G.~Nogues, L.~Constantin, P.~Durand, A.~Amy-Klein,
  A.~Van~Lerberghe, and C.~Chardonnet.
\newblock \emph{{CO}$_2$ laser stabilization to 0.1-{Hz} level using external
  electrooptic modulation}.
\newblock \emph{IEEE J. Quant. Electron.}, \textbf{33}(8):1282--1287 (Aug
  1997).
\newblock ISSN 0018-9197.
\newblock \doilink{10.1109/3.605548}.

\bibitem{Chen2014}
Q.-F. Chen, A.~Nevsky, M.~Cardace, S.~Schiller, T.~Legero, S.~H\"afner, A.~Uhde,
  and U.~Sterr.
\newblock \emph{A compact, robust, and transportable ultra-stable laser with a
  fractional frequency instability of $1\times10^{-15}$}.
\newblock \emph{Rev. Sci. Instrum.}, \textbf{85}(11):113107 (2014).
\newblock \doilink{http://dx.doi.org/10.1063/1.4898334}.

\bibitem{Galli2013}
I.~Galli, M.~Siciliani~de Cumis, F.~Cappelli, S.~Bartalini, D.~Mazzotti,
  S.~Borri, A.~Montori, N.~Akikusa, M.~Yamanishi, G.~Giusfredi, P.~Cancio, and
  P.~De~Natale.
\newblock \emph{Comb-assisted subkilohertz linewidth quantum cascade laser for
  high-precision mid-infrared spectroscopy}.
\newblock \emph{Appl. Phys. Lett.}, \textbf{102}(12):121117 (2013).
\newblock \doilink{10.1063/1.4799284}.

\bibitem{Hansen2013}
M.~G. Hansen, I.~Ernsting, S.~V. Vasilyev, A.~Grisard, E.~Lallier,
  B.~G\'{e}rard, and S.~Schiller.
\newblock \emph{Robust, frequency-stable and accurate mid-{IR} laser
  spectrometer based on frequency comb metrology of quantum cascade lasers
  up-converted in orientation-patterned {GaAs}}.
\newblock \emph{Opt. Express}, \textbf{21}(22):27043--27056 (Nov 2013).
\newblock \doilink{10.1364/OE.21.027043}.

\bibitem{Meyer1995}
B.~Meyer, S.~Saupe, M.~Wappelhorst, T.~George, F.~K\"uhnemann, M.~Schneider,
  M.~Havenith, W.~Urban, and J.~Legrand.
\newblock \emph{{CO}-laser side-band spectrometer: Sub-Doppler heterodyne
  frequency measurements around 5 $\mu$m}.
\newblock \emph{Appl. Phys. B}, \textbf{61}(2):169--173 (1995).
\newblock ISSN 0946-2171.
\newblock \doilink{10.1007/BF01090939}.

\bibitem{Mills2012}
A.~A. Mills, D.~Gatti, J.~Jiang, C.~Mohr, W.~Mefford, L.~Gianfrani, M.~Fermann,
  I.~Hartl, and M.~Marangoni.
\newblock \emph{Coherent phase lock of a 9 $\mu$m quantum cascade laser to a 2
  $\mu$m thulium optical frequency comb}.
\newblock \emph{Opt. Lett.}, \textbf{37}(19):4083--4085 (Oct 2012).
\newblock \doilink{10.1364/OL.37.004083}.

\bibitem{Sow2014}
P.~L.~T. Sow, S.~Mejri, S.~K. Tokunaga, O.~Lopez, A.~Goncharov, B.~Argence,
  C.~Chardonnet, A.~Amy-Klein, C.~Daussy, and B.~Darqui\'e.
\newblock \emph{A widely tunable 10-$\mu$m quantum cascade laser phase-locked
  to a state-of-the-art mid-infrared reference for precision molecular
  spectroscopy}.
\newblock \emph{Appl. Phys. Lett.}, \textbf{104}(26):264101 (2014).
\newblock \doilink{10.1063/1.4886120}.

\bibitem{Taubman2004}
M.~S. Taubman, T.~L. Myers, B.~D. Cannon, and R.~M. Williams.
\newblock \emph{Stabilization, injection and control of quantum cascade lasers,
  and their application to chemical sensing in the infrared}.
\newblock \emph{Spectrochim. Acta Mol.}, \textbf{60}(14):3457--3468 (2004).
\newblock ISSN 1386-1425.
\newblock \doilink{10.1016/j.saa.2003.12.057}.

\bibitem{Tombez2012}
L.~Tombez, S.~Schilt, J.~Di~Francesco, T.~F\"uhrer, B.~Rein, T.~Walther,
  G.~Di~Domenico, D.~Hofstetter, and P.~Thomann.
\newblock \emph{Linewidth of a quantum-cascade laser assessed from its
  frequency noise spectrum and impact of the current driver}.
\newblock \emph{Appl. Phys. B}, \textbf{109}(3):407--414 (2012).
\newblock ISSN 0946-2171.
\newblock \doilink{10.1007/s00340-012-5005-x}.

\bibitem{Wang2011}
L.~Wang, Z.~Cao, H.~Wang, H.~Zhao, W.~Gao, Y.~Yuan, W.~Chen, W.~Zhang, Y.~Wang,
  and X.~Gao.
\newblock \emph{A widely tunable (5--12.5 $\mu$m) continuous-wave mid-infrared
  laser spectrometer based on difference frequency generation in {AgGaS$_2$}}.
\newblock \emph{Opt. Commun.}, \textbf{284}(1):358--362 (2011).
\newblock ISSN 0030-4018.
\newblock \doilink{10.1016/j.optcom.2010.08.057}.

\end{thebibliography}

\end{document}